\documentclass[twocolumn,aps,prb,superscriptaddress]{revtex4-2}

\usepackage{graphicx}
\usepackage{multirow}
\usepackage{bm}
\usepackage{color}
\usepackage{amsmath}
\usepackage{ulem}
\usepackage{soul}
\usepackage{balance}

\begin{document}

\title[Shuaishuai Luo, et al]{Direction dependent switching of carrier-type enabled by Fermi surface geometry}

\author{Shuaishuai Luo}

\thanks{These authors contributed equally to this work.}
\affiliation{Center for Correlated Matter and School of Physics, Zhejiang University, Hangzhou 310058, China}

\author{Feng Du}

\thanks{These authors contributed equally to this work.}
\affiliation{Center for Correlated Matter and School of Physics, Zhejiang University, Hangzhou 310058, China}

\author{Dajun Su}

\affiliation{Center for Correlated Matter and School of Physics, Zhejiang University, Hangzhou 310058, China}

\author{Yongjun Zhang}

\affiliation{Institute for Advanced Materials, Hubei Normal University, Huangshi 435002, China}

\author{Jiawen Zhang}

\affiliation{Center for Correlated Matter and School of Physics, Zhejiang University, Hangzhou 310058, China}

\author{Jiacheng Xu}

\affiliation{Center for Correlated Matter and School of Physics, Zhejiang University, Hangzhou 310058, China}

\author{Yuxin Chen}

\affiliation{Center for Correlated Matter and School of Physics, Zhejiang University, Hangzhou 310058, China}

\author{Chao Cao}
\email[]{ccao@zju.edu.cn}

\affiliation{Center for Correlated Matter and School of Physics, Zhejiang University, Hangzhou 310058, China}

\author{Michael Smidman}
\email[]{msmidman@zju.edu.cn}

\affiliation{Center for Correlated Matter and School of Physics, Zhejiang University, Hangzhou 310058, China}

\author{Frank Steglich}
\affiliation{Center for Correlated Matter and School of Physics, Zhejiang University, Hangzhou 310058, China}
\affiliation{Max Planck Institute for Chemical Physics of Solids, 01187 Dresden, Germany}

\author{Huiqiu Yuan}
\email[]{hqyuan@zju.edu.cn}

\affiliation{Center for Correlated Matter and School of Physics, Zhejiang University, Hangzhou 310058, China}
\affiliation  {State Key Laboratory of Silicon Materials, Zhejiang University, Hangzhou 310058, China}
\affiliation  {Collaborative Innovation Center of Advanced Microstructures, Nanjing 210093, China}

\date{\today}

\begin{abstract}
	While charge carriers can typically be designated as either electron- or hole- type, depending on the sign of the Hall coefficient, some materials defy this straightforward classification. Here we find that LaRh$_6$Ge$_4$ goes beyond this dichotomy, where the Hall resistivity is electron-like for magnetic fields along the $c$-axis but hole-like in the basal plane. Together with first-principles calculations, we show that this direction-dependent switching of the carrier type arises within a single band, where the special geometry leads to charge carriers on the same Fermi surface orbiting as electrons along some directions, but holes along others. The relationship between the Fermi surface geometry and occurrence of a Hall sign reversal is further generalized by considering tight-binding model calculations, which show that this type of Fermi surface corresponds to a more robust means of realizing this phenomenon, suggesting an important route for tailoring direction dependent properties for advanced electronic device applications.
\end{abstract}

\maketitle

\section{Introduction}
The Hall effect corresponds to the deflection of charge carriers  in a magnetic field perpendicular to the current direction, and has an intimate link to the electronic structure of a material. As only electrons near the Fermi energy will redistribute under external perturbations, the Hall effect, as well many other electronic properties, in a metal is largely determined by the nature of the Fermi surface \cite{Kaganov1979,Kaganov2002}. This phenomenon has received tremendous attention in a variety of contexts, including the anomalous Hall effect \cite{nagaosa2010anomalous}, topological Hall effect \cite{neubauer2009topological}, quantum Hall effect \cite{zhang2005experimental} and quantum anomalous Hall effect \cite{chang2013experimental}. On the other hand, the ordinary Hall effect has served as a powerful tool for probing the nature of the charge carriers \cite{nHall2009,Badoux2016,nHall2010}, where the sign of the Hall coefficient $R_H$ indicates the type of charge carrier (electron or hole), while its magnitude corresponds to the carrier density. 
This simple picture based on the Drude model is readily applicable in many scenarios  but can breakdown in more complicated cases, 
where instead the transport properties can be calculated from the Boltzmann equation, provided there is complete information about the band structure and relaxation times \cite{jones1934theory,ziman2001electrons}. Furthermore, in multiband materials electron and hole carriers can coexist, with each type being associated with different bands \cite{Zhu2015,Luo2015,xu2017origin}.

\begin{figure*}[tb]
	\includegraphics[scale=0.44]{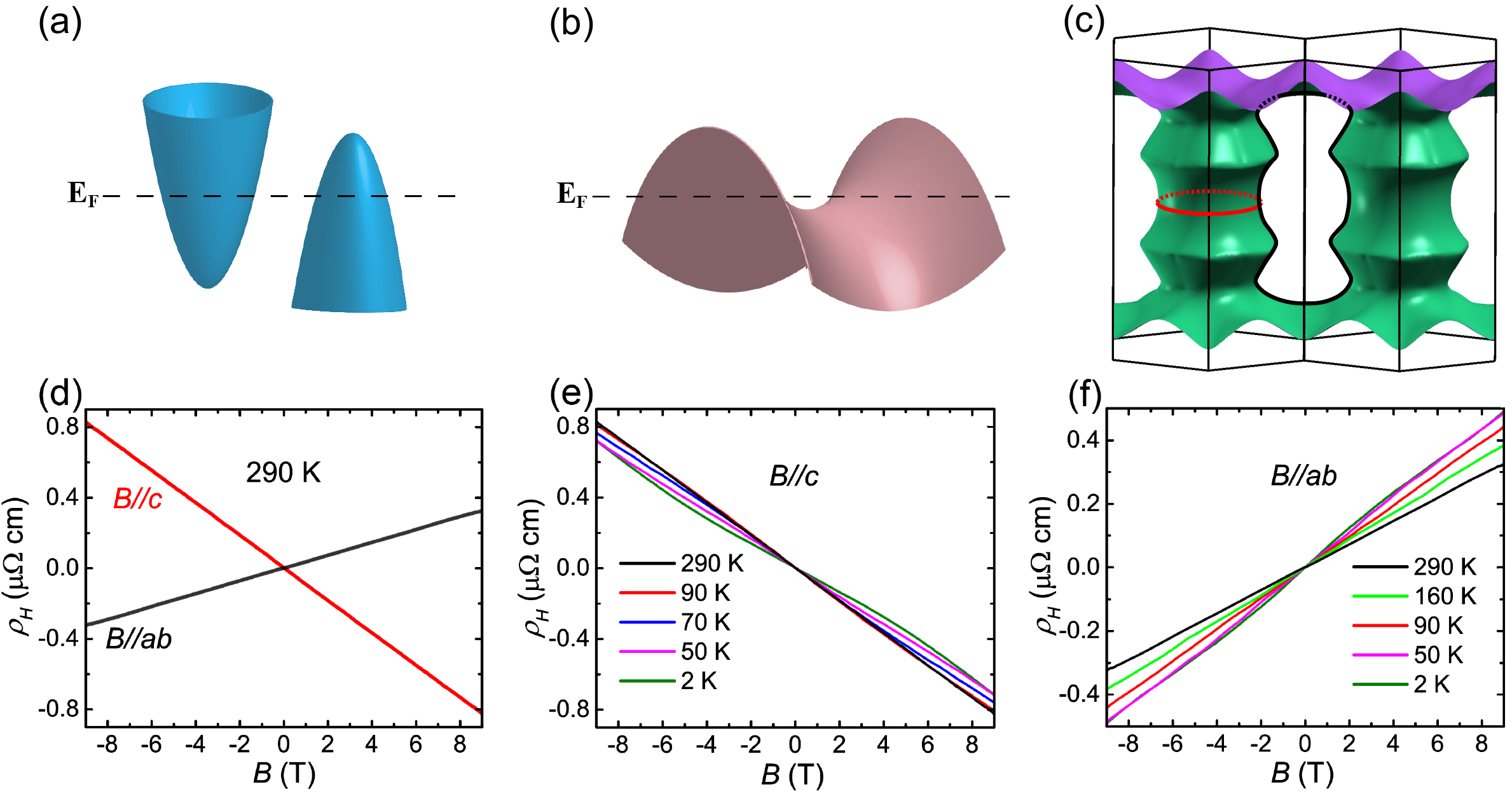}%
	\caption{\label{fig:scenarios} Overview of the Hall effect sign reversal (HSR) in  LaRh$_6$Ge$_4$. Two-possible scenarios for realizing the HSR are (a) a multiband scenario with both hole and electron bands, and (b) a single band with a saddle point.  (c) Calculated Fermi surface corresponding to the  $\beta$ band of LaRh$_6$Ge$_4$ across two neighboring Brillouin zones. The red circle and black bow-knot highlight the closed electron orbit associated with $B\parallel c$ and the closed hole orbit associated with  $B\parallel ab$, respectively. (d) Demonstration of HSR in LaRh$_6$Ge$_4$ at room temperature, where the Hall resistivity has a positive slope for magnetic fields applied  in the $ab$-plane, and a negative slope along the $c$ axis. Field dependence of the Hall resistivity at various temperatures between 2 and 290~K for (e) $B\parallel c$,  and (f) $B\parallel ab$.}
\end{figure*}

A striking example that goes beyond the aforementioned dichotomy between holes and electrons  is systems where the Hall coefficient changes sign upon adjusting the magnetic field (and current) direction, namely there is a Hall effect sign reversal (HSR). Such a phenomenon has been observed in a handful of systems \cite{mondal2020extremely,araki2014charge,eltsev2002anisotropic,he2019fermi}, and in most of these examples  the behaviors are accounted for by the multiband scenario illustrated in Fig.~\ref{fig:scenarios}(a), where there are coexistent electron and hole pockets with anisotropic mobilities \cite{SM}. Alternatively, HSR can also arise from a single band with a saddle point dispersion (Fig.~\ref{fig:scenarios}(b)), such that charge carriers act as electrons along some directions and holes along others \cite{SM}. Such a situation is more unusual compared to the  multiband scenario, whereby the charge carriers cannot be classified as purely electron- or hole-type \cite{Skinner2019}. In NaSn$_2$As$_2$, there  are closed hole-type orbits for out-of-plane magnetic fields, but open orbits for in-plane fields, and the electron-type behavior of the latter is due to the balance between concave and convex portions of the Fermi surface \cite{he2019fermi}. A very different scenario for realizing HSR in a single band is shown by the Fermi surface in Fig.~\ref{fig:scenarios}(c), for which both out-of-plane and in-plane magnetic fields will induce closed orbits, where the former encircles occupied states and hence corresponds to electron-type (red circle), while the latter encloses unoccupied states and therefore describes a hole-type orbit (black bow-knot).

Here we find that  LaRh$_6$Ge$_4$ exhibits HSR enabled by the peculiar Fermi surface geometry in Fig.~\ref{fig:scenarios}(c). Unlike its Ce analog which displays significant electronic correlations and ferromagnetic quantum criticality \cite{Vowinkel2012,Shen2020}, LaRh$_6$Ge$_4$ is a weakly correlated nonmagnetic material \cite{SM,vosswinkel2013new}, and therefore  the anomalous Hall effect is absent, allowing for the ordinary contributions to the Hall resistivity from the charge carriers to be readily examined. First principles calculations suggest that the transport properties are dominated by the spin-orbit split $\beta$ (and $\beta'$) bands, which form the aforementioned Fermi surface displayed in Fig.~\ref{fig:scenarios}(c), and hence LaRh$_6$Ge$_4$ is a prime candidate for HSR arising from a single band. 

\section{Methods}
\subsection{Experimental details}

Single crystals of LaRh$_6$Ge$_4$ were grown using a Bi flux method \cite{vosswinkel2013new}. Hall resistivity measurements were performed in Quantum Design Physical Property Measurement System (PPMS) using the four-contact method. To perform the angular dependent measurements, the samples were polished in order to obtain $I-V$ planes at different angles $\theta$ to the $ab$ plane. The corresponding plane indices ($h,k,l$) were then identified through the Laue patterns, and the angles $\theta$ were calculated based on the experimentally determined hexagonal crystal structure and lattice parameters in Ref. \cite{vosswinkel2013new}: 
\begin{equation}
\theta= {\rm{arccos}}\frac{(h\bm{a}^*+k\bm{b}^*+l\bm{c}^*)\cdot\bm{c}^*}{\lvert h\bm{a}^*+k\bm{b}^*+l\bm{c}^*\rvert\cdot\lvert\bm{c}^*\lvert}
\label{theta}
\end{equation}
where $\bm{a}^*,\bm{b}^*,\bm{c}^*$  are basis vectors in  reciprocal space. The magnetic field was always perpendicular to the $I-V$ planes in all the measurements.

\subsection{Calculations of band structure, Fermi surfaces and Hall coefficients}

First principles calculations were performed using the plane-wave projected augmented wave method as implemented in the VASP code. The Perdew, Burke and Ernzerhoff parameterization (PBE) of the general gradient approximation (GGA) was used for the exchange-correlation functionals \cite{PBE1996}. For calculations of the Fermi surfaces in LaRh$_6$Ge$_4$, band structures from VASP were fitted to a tight-binding Hamiltonian with 108 atomic orbitals including La-5$d/4f$, Rh-4$d$ and Ge-4$p$, using the maximally projected Wannier function
method \cite{PhysRevB.71.125119}. The resulting Wannier-orbital-based Hamiltonian was symmetrized using full crystal symmetry \cite{Zhi2022}, and
was used to calculate the Fermi surfaces by interpolating the band structure to a 100$\times$100$\times$100 dense
k-mesh. While for the single-band tight-binding model, the Fermi surfaces can be readily calculated
according to the analytical band dispersion.

Hall coefficients were calculated in the framework of  Boltzmann transport theory, in which current density can be expressed up to first order in $B$ as \cite{ziman2001electrons},
\begin{equation}
	J_\alpha=\sigma_{\alpha\beta}E_\beta+\sigma_{\alpha\beta\gamma}E_\beta B_\gamma+...
	\label{eq:e1}.
\end{equation}

\noindent For a multi-band system, assuming the relaxation time is isotropic for each band, we have:

\begin{equation}
	\sigma_{\alpha\beta}=e^2\sum_n \int \frac{d^3\mathbf{k}}{(2\pi)^3} \tau_{n\mathbf{k}} \left(-\frac{\partial f_{n\mathbf{k}}}{\partial \varepsilon}\right) v_{n\mathbf{k}}^{\alpha} v_{n\mathbf{k}}^{\beta} \label{eq:e2}.
\end{equation}

\begin{equation}
	\sigma_{\alpha\beta\gamma}=-e^3\sum_{n,\beta',\alpha'} \epsilon_{\gamma\beta'\alpha'}\int \frac{d^3\mathbf{k}}{(2\pi)^3} \tau_{n\mathbf{k}}^2 \left(-\frac{\partial f_{n\mathbf{k}}}{\partial \varepsilon}\right) \Xi^{\alpha\alpha'\beta\beta'}_{n\mathbf{k}}
	\label{eq:e4}
\end{equation}

\begin{equation}
	\Xi^{\alpha\alpha'\beta\beta'}_{n\mathbf{k}}=v_{n\mathbf{k}}^{\alpha} v_{n\mathbf{k}}^{\alpha'} \left[M^{-1}\right]_{n\mathbf{k}}^{\beta\beta'}
	\label{eq:e3}
\end{equation}

\noindent where $(\alpha, \beta, \gamma)$ is a permutation of $(x, y, z)$, $\epsilon_{\alpha\beta\gamma}$ is the antisymmetric tensor, $n$ is the band index, $f_{n\mathbf{k}}=f(\epsilon_{n\mathbf{k}})$ is the Fermi-Dirac function, $\epsilon_{n\mathbf{k}}$ is the $n$-th band energy at $\mathbf{k}$ (Fermi energy $\epsilon_F=0$). The group velocity $v_{n\mathbf{k}}^\alpha$ and inverse effective mass tensor $\left[M^{-1}\right]_{n\mathbf{k}}^{\alpha\beta}$ are defined as:

\begin{equation}
	v_{n\mathbf{k}}^\alpha=\frac{1}{\hbar}\frac{\partial \epsilon_{n\mathbf{k}}}{\partial k_{\alpha}}
\end{equation}

\begin{equation} \left[M^{-1}\right]_{n\mathbf{k}}^{\alpha\beta}=\frac{1}{\hbar^2}\frac{\partial^2 \epsilon_{n\mathbf{k}}}{\partial k_{\alpha}\partial k_{\beta}}
\end{equation}

\noindent For calculations based on first-principles Wannier Hamiltonians, the calculation of these quantities follows Ref. \cite{PhysRevB.75.195121}; while for the tight-binding model, the group velocity and inverse effective mass tensor are analytically calculated according to the band dispersion, where the integrals are replaced by summations over a 100$\times$100$\times$100 dense mesh.

\noindent The Hall coefficient is then defined as:
\begin{equation}
	R^{H}_{\alpha\beta\gamma}=\frac{E_{\beta}}{j_{\alpha}B_{\gamma}}=\sum_{\alpha'\beta'}\left[\sigma^{-1}\right]_{\alpha'\beta}\sigma_{\alpha'\beta'\gamma}\left[\sigma^{-1}\right]_{\alpha\beta'}
	\label{eq:e5}.
\end{equation}

\section{Results and Discussion}
The presence of HSR in LaRh$_6$Ge$_4$ is demonstrated in Fig.~\ref{fig:scenarios}(d), where the Hall coefficient $R_H$ is positive when the magnetic field is applied in the $ab$-plane, but negative for fields along the $c$-axis. At this elevated temperature, the Hall resistivity has a linear field dependence for both directions, indicating the dominance of a single carrier type. As shown in Figs.~\ref{fig:scenarios}(e) and (f), upon decreasing the temperature there are slight deviations from linear behavior, suggesting small contributions from other carriers, but the HSR remains robust down to at least 2~K. These behaviors in LaRh$_6$Ge$_4$ are reproducible in different samples  \cite{SM}.

\begin{figure}[tb]
	\includegraphics[scale=0.48]{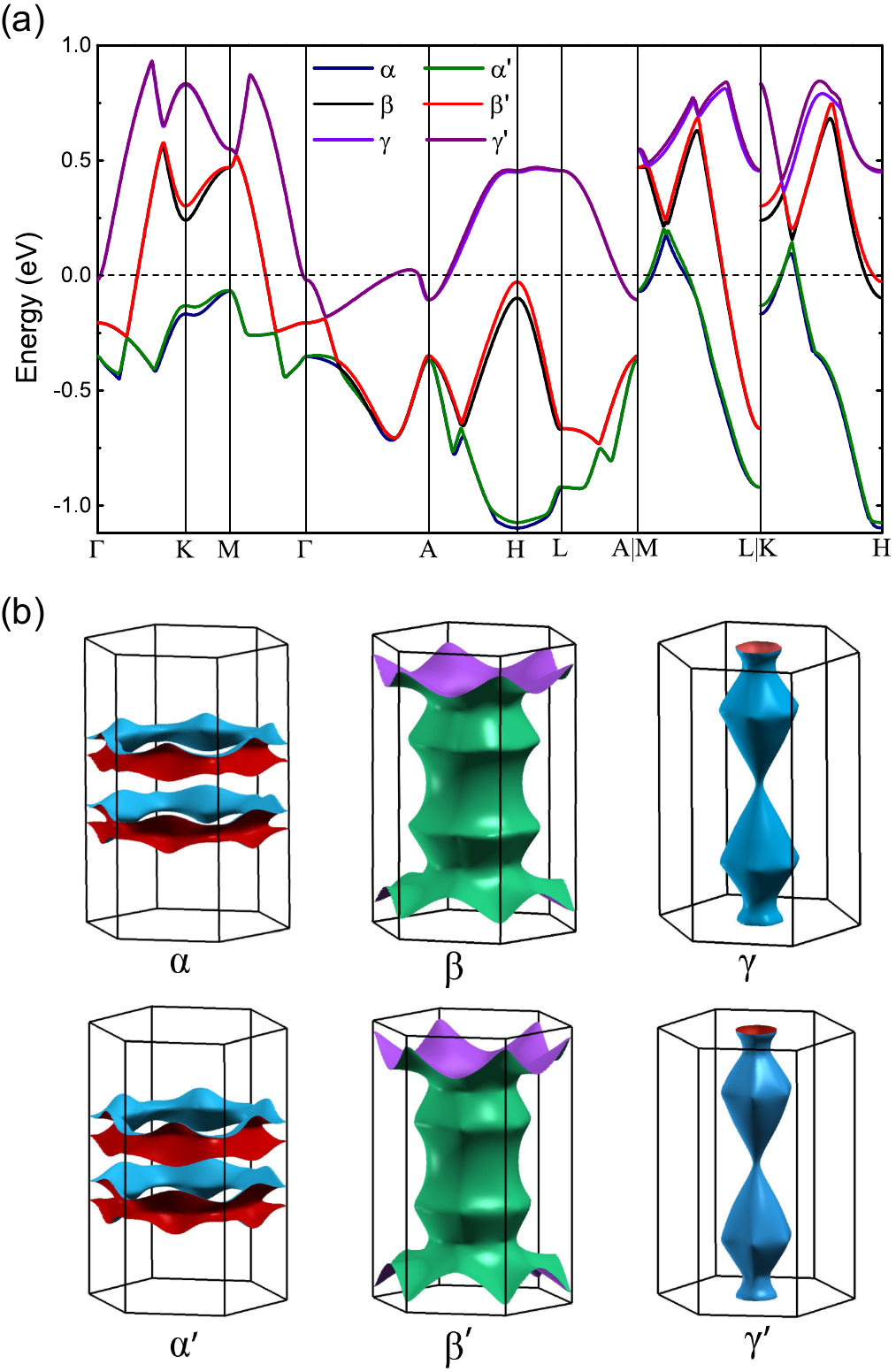}
	\caption{\label{fig:band} Electronic structure of LaRh$_6$Ge$_4$. (a) Band structure of LaRh$_6$Ge$_4$ obtained from DFT calculations. (b) Fermi surfaces corresponding to the $\alpha$, $\beta$ and $\gamma$ bands together with their SOC-split counterparts, where the larger $\beta$ surface corresponds to that of Fig.~\ref{fig:scenarios}(c).}
\end{figure}

\begin{figure*}[t]
	\includegraphics[scale=0.67]{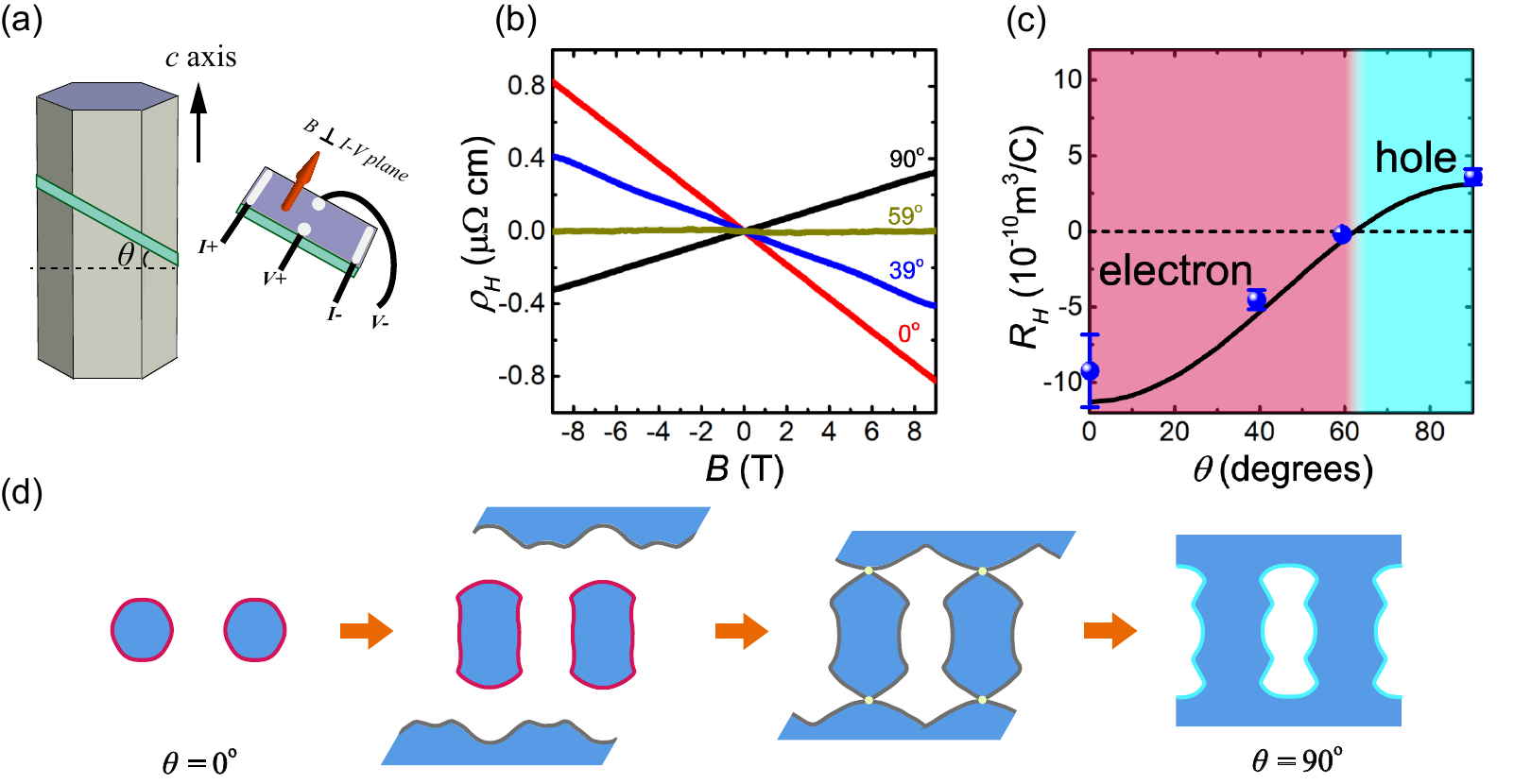}%
	\caption{\label{fig:angular} (a) Schematic diagram for the angular dependent Hall resistivity measurements, where $\theta$ is the angle between the $ab$-plane and the $I$-$V$ plane. (b) Field-dependence of the room temperature Hall resistivity for  different $\theta$, where for $\theta=59^{\circ}$ the Hall coefficient $R_H\approx 0$. (c) Evolution of the room temperature Hall coefficient as a function of $\theta$. Blue symbols are from experiments, while the black solid line shows the calculated $R_H$ from theoretical calculations. The calculated orbit types are shown by the color plot, where red and cyan  represent electron and hole type respectively. (d) Schematic diagram illustrating how the orbit-type evolves with $\theta$ from electron-type closed orbits enclosing occupied states at $\theta=0^{\circ}$ (red line), to hole-type closed orbits enclosing unoccupied states at $\theta=90^{\circ}$ (cyan line).}
\end{figure*}

Figure~\ref{fig:band}(a) displays the results of band structure calculations for LaRh$_6$Ge$_4$ based on density-functional theory (DFT) with spin orbit coupling (SOC) taken into account. Due to the lack of inversion symmetry, the SOC lifts the spin degeneracy of the electron bands. Figure~\ref{fig:band}(b) shows the corresponding Fermi surfaces $\alpha, \beta, \gamma$ and their SOC-split counterparts $\alpha', \beta', \gamma'$. The surfaces belonging to each pair of surfaces show almost identical features. The calculations are consistent with previous studies of LaRh$_6$Ge$_4$ and similar to isostructural CeRh$_6$Ge$_4$ \cite{Guo2018,WANG2021}. The Fermi level is situated in the middle of the  $\beta$ and $\beta'$  bands, where $48.0\%$  of the $\beta$ states are occupied, while $\alpha$ is almost full ($93.7\%$ occupation), and $\gamma$ is almost empty ($6.6\%$ occupied). Consequently the much larger $\beta$ and $\beta'$ Fermi surfaces, which exhibit the geometry displayed in Fig.~\ref{fig:scenarios}(c), are anticipated to make the dominant contribution  to the transport properties. The calculated $R_H$ values based on the band structure  well reproduce the observed HSR, with calculated room temperature values of $-11.26\times10^{-10}$~m$^3$/C and $3.11\times10^{-10}$~m$^3$/C for $B\parallel c$ and $B\parallel ab$, respectively, which are consistent with the respective experimental values of $-9.22\times10^{-10}$~m$^3$/C and $3.63\times10^{-10}$~m$^3$/C.

In order to confirm that this HSR can be understood by considering only one Fermi surface pocket, the Hall resistivity for different field angles were measured with the configuration illustrated in Fig.~\ref{fig:angular}(a), where the samples were  polished in order to ensure that different crystal planes correspond to the $I$-$V$ plane, which is perpendicular to the magnetic field. The angle $\theta$ between the measured $I$-$V$ plane and the $ab$ plane was identified using the x-ray Laue method, and the results are shown in Fig.~\ref{fig:angular}(b).

At $\theta=0^{\circ}$ ($B\parallel c$), $R_H$ has a large negative value, which increases with increasing $\theta$, and at $\theta=59^{\circ}$ $\rho_H$ is flat with  $R_H\approx0$, while $R_H$ is positive at larger $\theta$. The colormap in Figure~\ref{fig:angular}(c) shows the angle-dependence of the orbit-type for the $\beta$ surface, based on the occupancy of the electronic states in DFT calculations, where the red and cyan region represent electron- and hole-type respectively. The blue symbols and black solid line represent $R_H$ from experiments and calculations, respectively. It can be seen that the orbit changes from electron to hole type between 60$^{\circ}$ and 65$^{\circ}$, very close to the angle where $R_H\approx0$ in experiments. This demonstrates that the HSR can indeed be understood as arising from the single band scenario of Fig.~\ref{fig:scenarios}.

How the nature of the cyclotron orbits evolves upon rotating the magnetic field from the $c$-axis to the $ab$-plane is illustrated schematically in Fig.~\ref{fig:angular}(d). At $\theta=0^{\circ}$, there are nearly circular orbits enclosing occupied states, corresponding to electron orbits. At intermediate angles, the electron orbits become elongated and additional  open orbits emerge, and at higher angles these previously separated Fermi surface contours intersect at a Lifshitz-like transition. Beyond this point, the resulting closed orbits correspond to hole orbits enclosing unoccupied states. As $R_H$ at the two extremal angles have different signs, there should be a critical angle at which $R_H=0$. Intuitively this can be understood as arising from the Fermi surface curvature, whereby at intermediate angles segments with positive and negative curvature contribute with different signs and the total value depends on competition between them. While the critical angle does not necessarily coincide with the Lifshitz-like transition, the Fermi surface curvature changes rapidly within its vicinity, and therefore the crossover between hole and electron type behaviors occurs in this region.

\begin{figure}[t]
	\includegraphics[scale=0.48]{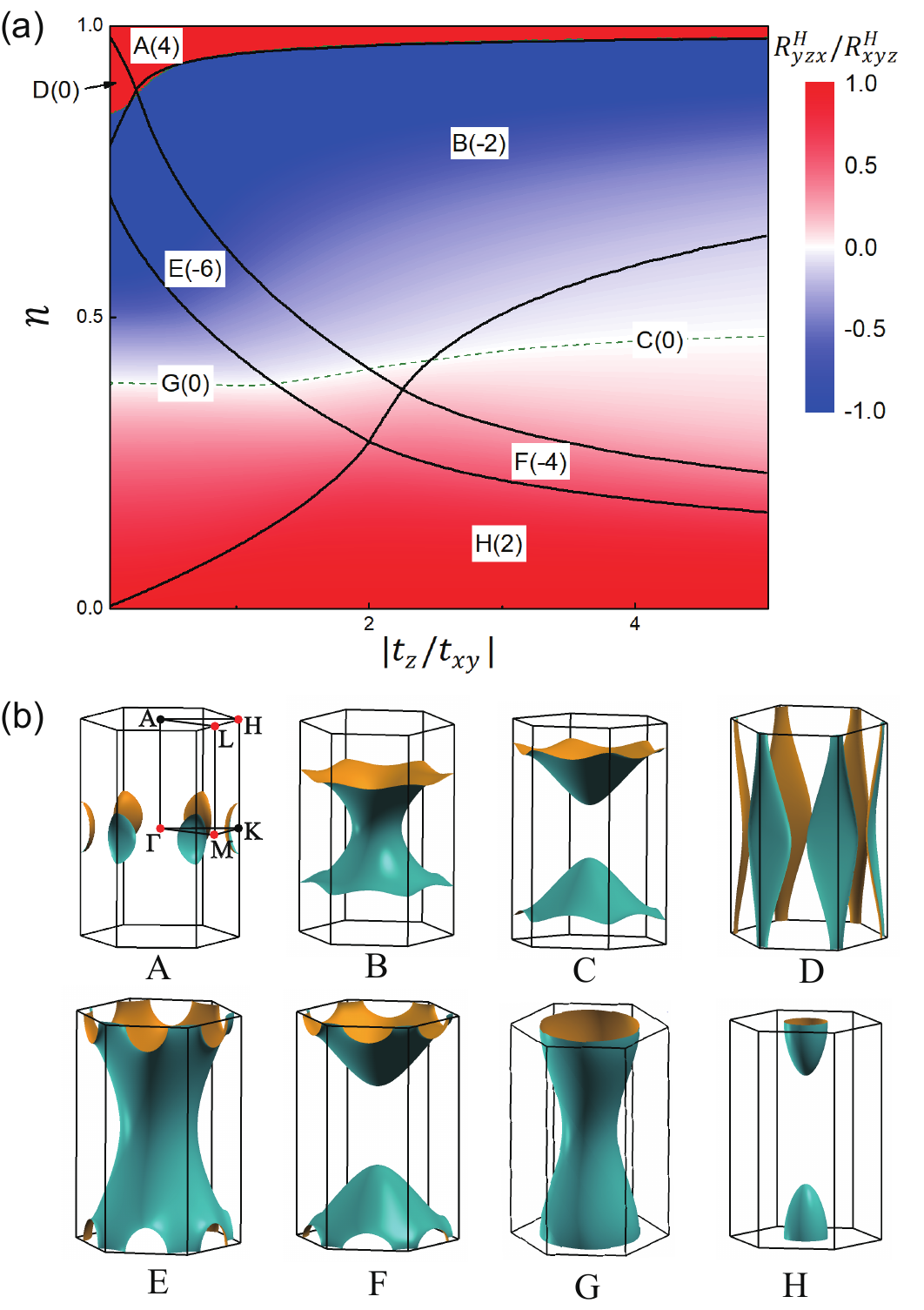}%
	\caption{\label{fig:model} HSR from a tight-binding model on a hexagonal lattice. (a) Phase  diagram as a function of filling factor $n$ and ratio of the interlayer to intralayer hopping parameters $\lvert t_z/t_{xy}\rvert$ (with $t_{xy}>0$ and $t_z<0$) for the hexagonal lattice tight-binding model described in the text. The solid lines correspond to Lifshitz transitions between different Fermi surface geometries labelled A-H, while the numbers in the parentheses correspond to the modified Euler characteristic  $\chi_M^\star$. The color plot corresponds to the ratio between the Hall coefficients for the two perpendicular directions, and therefore blue regions where this ratio is negative exhibit HSR. (b) Plots of the different Fermi surfaces A-H obtained from the tight-binding model.}
\end{figure}

\begin{figure}[h]
	\includegraphics[scale=0.34]{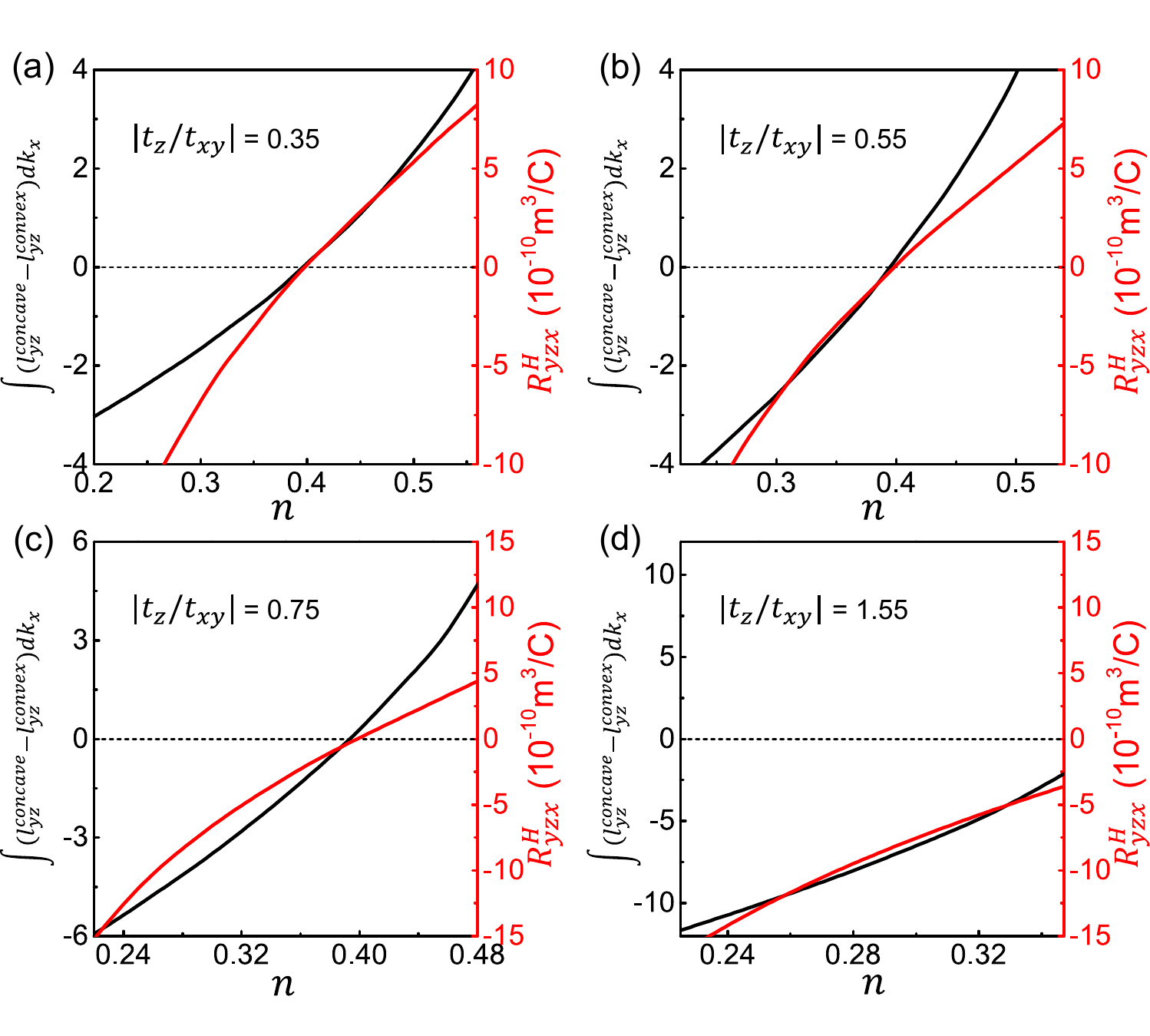}
	\caption{\label{fig:condition} The integral of the difference between the projected scattering path lengths of the concave and convex portions, together with the calculated $R_{yzx}^H$, for the hyperboloid Fermi surface G from the tight-binding model calculations. In (a), (b), (c) there is a sign change of the integral and the Hall coefficient at almost exactly the same filling, while at larger $\lvert t_z/t_{xy}\rvert$ in (d), both quantities are consistently negative for surface G.}
\end{figure}

To gain a more systematic understanding of the scenarios giving rise to HSR within one band, we considered a single-band tight-binding model on a hexagonal lattice with only nearest-neighbor hopping, yielding the following band dispersion,
\begin{equation}
\epsilon_{\mathbf{k}}=\epsilon_0-2t_z\cos k_c
-2t_{xy}\left[\cos k_1+\cos k_2+\cos (k_1+k_2)\right].
\label{equation1}
\end{equation}

\noindent where $(k_1, k_2, k_c)$ are the  $\mathbf{k}$-point coordinates in units of reciprocal lattice vectors. This model gives rise to a set of saddle points at high symmetry points, namely $\Gamma$, $L$, $H$, and $M$ ($A$, $L$, $M$, and $K$) if the inter-plane hopping $t_z$ has a different (the same) sign as the intra-plane hopping $t_{xy}$, leading to a series of Lifshitz transitions upon changing the filling factor $n$. Figure~\ref{fig:model}(a) displays the resulting phase diagram for $n$ versus $\lvert t_z/t_{xy}\rvert$ (with $t_{xy}>0$ and $t_z<0$), where the regions enclosed by the solid lines each represent a different Fermi surface geometry, which are displayed in Fig.~\ref{fig:model}(b). The Hall coefficients were calculated, and the color plot corresponds to $R^H_{yzx}/R^H_{xyz}$ (subscripts indicate respective directions of current, voltage and field), such that the blue regions ($R^H_{yzx}/R^H_{xyz}<0$) exhibit a HSR but red areas do not.

Upon increasing $n$ for small $\lvert t_z/t_{xy}\rvert$, there is a Lifshitz transition when the Fermi level crosses the saddle point at $\Gamma$, at which there is a change from the ellipsoid Fermi surface labelled H, to the open hyperboloid G. The latter has closed electron orbits within the hexagonal plane, while it is open along $\Gamma-A$ with both convex and concave portions. According to the geometric representation of 2D metal proposed in Ref.~\onlinecite{ong1991geometric}, the Hall contribution can be readily obtained by mapping the Fermi surface from $\mathbf{k}$ space to the space of scattering path length ($\bm{l}$ space) and finding the total ``Stokes" area. Thus, the Hall sign change can happen for an open orbit only when the scattering path length of convcave portions is larger than that of the convex portion \cite{ong1991geometric,SM}. A 3D generalization of the geometric representation analysis requires integration over the third dimension, but the integral of the projected scattering path length $\int l_{\perp} dk_\parallel$ still serves as an approximate criterion. Figure~\ref{fig:condition} shows the integral of the difference between the projected scattering path lengths $\int(l_{yz}^{concave}-l_{yz}^{convex}) dk_x$, together with the calculated Hall coefficient $R_{yzx}^H$, within the region corresponding to the hyperboloid Fermi surface G, for different values of $\lvert t_z/t_{xy}\rvert$. The sign change of the integral coincides with where $R_{yzx}^H$ changes sign for small $\lvert t_z/t_{xy}\rvert$, while for large $\lvert t_z/t_{xy}\rvert$, both the integral and $R_{yzx}^H$ remain negative up to the transition to a different geometry. These suggest that the sign change of the Hall coefficient for the G Fermi surface is indeed due to the competition between the concave and convex portions, which corresponds to the scenario in NaSn$_2$As$_2$ \cite{he2019fermi}.

On the other hand, further increasing $n$ can lead to the Fermi level moving across the two saddle points at $L$ and $H$, giving rise to Fermi surface B, which corresponds to the $\beta$ band of LaRh$_6$Ge$_4$. This represents a very different scenario, whereby there are closed orbits along perpendicular directions corresponding to different orbit-types, and as such this exhibits a much more robust HSR covering nearly all the parameter space for this Fermi surface. Although the balance between the concave and convex portions plays a role at intermediate angles where open orbits emerge (as depicted in Fig.~\ref{fig:angular}(d), especially around the critical angle where such a balance leads to a vanishing Hall coefficient), the HSR between the two extremal angles is robust and moderate tuning of the local curvature of the Fermi surface only adjusts the exact position of the critical intermediate angle.

Furthermore, in analogy to the Euler characteristic used to classify the topology of surfaces, the topology of these Fermi surfaces can be characterized by a modified version $\chi_M^\star=\frac{1}{2\pi}\int KdS$, where $K$ denotes the Gaussian curvature \cite{tu2017differential,SM}. For Fermi surface B which corresponds to that of LaRh$_6$Ge$_4$, $\chi_M^\star$ has a  value of $-2$, while Fermi surface E which also exhibits HSR across most of the parameter space has a value of $-6$. On the other hand, Fermi surfaces G and C have $\chi_M^\star=0$, and have open orbits along some directions, and therefore the sign of the Hall coefficient for these depends on the balance between regions with positive and negative curvature, and as such the HSR occurs over a much more limited region for these surfaces. This suggests that the value of $\chi_M^\star$ could be an indicator of Fermi surfaces which can exhibit a more robust HSR.

\section{Summary}
To summarize, we find that the Hall coefficient of LaRh$_6$Ge$_4$ exhibits different signs depending on whether the magnetic field is applied parallel or perpendicular to the $c$-axis. By combining experimental measurements of the Hall resistivity at different field-angles with the results of band structure calculations, we demonstrate that this sign-reversal originates from one of the Fermi surfaces, on which the charge carriers move as electrons on some closed orbits, and holes along others. Moreover, from considering a tight-binding model we obtain a more unified picture of the range of scenarios giving rise to  HSR from a single band and the  mechanisms for its realization. Materials with HSR also have the potential to be exploited in devices, where the dual electron and hole nature of the charge carriers can satisfy the requirements of advanced electronics. The findings of a Fermi surface geometry that robustly exhibits HSR should be particularly beneficial for high-throughput searches for candidate materials that possess this and other functional material properties. In particular,  the manifestation of HSR on a single band can greatly simplify the calculation of relevant electronic and thermal quantities \cite{Skinner2019}. 
Furthermore, the existence of a `magic angle' at which the Hall coefficient disappears could have applications in sensors and related devices. Consequently, the experimental findings combined with model calculations revealing a robust manifestation of HSR at room temperature for certain Fermi surface geometries provides the means for realizing advanced functional materials exhibiting this phenomenon.

\section*{Acknowledgments}
This work was supported by the National Key R\&D Program of China (Grant No. 2022YFA1402200), the Key R\&D Program of Zhejiang Province, China (Grant No. 2021C01002),  the National Natural Science Foundation of China (Grants No. 12274364, No. 12222410, No. 11974306, No. 12034017, and No. 12174332), and the Zhejiang Provincial Natural Science Foundation of China (Grant No. LR22A040002).

\normalem
\balance
%

\end{document}